\documentstyle[graphicx,multicol,prb,aps]{revtex}

\title{
Similarity and Difference between Magnetic- and Nonmagnetic-Impurity\\ 
Effects in Spin-Peierls Cuprate CuGeO${}_3$
}

\author{ 
N. {Koide},${}^{1,}$\cite{Koide-address}
Y. {Uchiyama},${}^1$ T. {Hayashi},${}^{1,2}$ T. {Masuda},${}^1$ 
Y. {Sasago},${}^{1,}$\cite{Sasago-address}
K. {Uchinokura},${}^{1,}$\cite{Uchinokura-email}\\
K. {Manabe},$^{2,}$\cite{Manabe-address} 
and H. {Ishimoto}$^2$
}

\address{
${}^1$Department of Applied Physics, The University of Tokyo, 7-3-1 Hongo, 
Bunkyo-ku, Tokyo 113-8656, Japan\\
${}^2$Institute for Solid State Physics, The University of Tokyo, 
7-22-1 Roppongi, Minato-ku, Tokyo 106-8666, Japan
}

\date{
\today
}
\begin{document}
\maketitle

\begin{multicols}{2}[
\begin{abstract}
We have measured the magnetic susceptibility of single-crystal 
Cu$_{1-x}$Ni$_x$GeO$_3$ to study doping effects of the magnetic impurity 
(Ni${}^{2+}$ has $S=1$ spin) on the inorganic spin-Peierls 
(SP) material CuGeO$_3$.
We observed the disappearance of the SP transition 
and the abrupt increase of the N\'eel temperature from 2.5 
to 3.4 K at $x\sim$ 0.020,
which indicates the first-order phase transition between the 
dimerized-antiferromagnetic (D-AF) and the uniform-antiferromagnetic (U-AF) 
phases, as was discovered in Mg-doped (Mg${}^{2+}$ is nonmagnetic) system 
\lbrack T. Masuda {\it et al.}, Phys. Rev. Lett. {\bf 80}, 4566 (1998)\rbrack.
This indicates that this transition is universal for the doping to Cu site.
We also found another phase with the easy axis in the $a-c$ plane
in low $x$ region. 
\end{abstract}
]
\narrowtext

The magnetic properties of CuGeO${}_3$ have been extensively studied
since Hase, Terasaki and Uchinokura \cite{ref1} reported that this compound
is the first inorganic spin-Peierls (SP) material.
The effects of Zn substitution for Cu  
were investigated
by Hase {\it et al.} \cite{ref2} and the suppression of SP 
transition and the occurrence of another phase transition were reported.
It has been established that the antiferromagnetic (AF) transition takes
place below 5 K in Cu${}_{1-x}M_{x}$GeO${}_3$ ($M$ = %
Zn (Ref.~\onlinecite{ref3,ref4,ref5}), Ni (Ref.~\onlinecite{ref3,ref5}), %
Mn (Ref.~\onlinecite{ref3})) and CuGe$_{1-x}$Si${}_x$O${}_3$ 
(Ref.~\onlinecite{ref6}).

In particular Cu${}_{1-x}$Zn$_{x}$GeO${}_3$ and
CuGe$_{1-x}$Si${}_x$O${}_3$ have been investigated in detail
by means of magnetization measurements,\cite{ref2,ref4,ref6} %
neutron scattering measurements,\cite{ref6,neutron1,Sasago,Martin} %
and so on.
One of the most interesting phenomena observed in 
Zn- and Si-doped CuGeO$_3$ is
the coexistence of the long-range order of the lattice dimerization,
which is intrinsic to SP phase,
and the AF long-range order (AF-LRO).\cite{ref6,Sasago}
Theoretical investigations of Si-doped CuGeO$_3$ have recently 
suggested
that local strain around the doped Si
reduces the SP lattice dimerization,
so that AF-LRO appears.\cite{Fukuyama}
This theory indicated that both the lattice order parameter
and the spin order parameter can coexist,
while they have a large spatial variation;
the size of the ordered moments (lattice dimerization) becomes 
maximum (minimum) around the doping center, and these spatial 
envelopes are described by the elliptic functions.\cite{Fukuyama}

On the other hand, for Zn-doped CuGeO$_3$, the mechanism of the
AF-LRO is not yet clear.
To substitute Zn for Cu in CuGeO$_3$ may have a meaning
to cut the $S = 1/2$ chains by $S = 0$ ions as well as  
to disturb the lattice systems.
Recently Masuda {\it et al.} discovered that there are two AF phases
in Cu${}_{1-x}$Mg${}_x$GeO${}_3$ (Mg${}^{2+}$ ion is nonmagnetic 
impurity as Zn${}^{2+}$ ion).\cite{masuda}
Both of the phases have AF-LRO,
i.e.,  they are the dimerized-antiferromagnetic 
(D-AF) phase (small $x$ region) and the 
uniform-antiferromagnetic (U-AF) phase (large $x$ region)
and there is a first-order phase transition between them 
at $x\sim$ 0.023.\cite{masuda}
In the U-AF phase, the lattice dimerization is absent and the state is 
a conventional AF state. 
On the other hand for $x<x_c$ the lattice is dimerized 
below the SP transition temperature ($T_{SP}$) 
and even below the N\'eel temperature ($T_N$), i.e.,
both of the order parameters coexist in the D-AF phase.

Here we must determine whether this transition is characteristic  
solely to Mg-doped CuGeO${}_3$ or is universal at least for the 
substitution for Cu${}^{2+}$.
For this purpose it is necessary to examine single-crystal 
Cu${}_{1-x}M_{x}$GeO${}_3$
in detail, where $M$ is a magnetic impurity.
One of the best candidates of magnetic impurities is Ni,
because Ni$^{2+}$ ion has $S = 1$ (Ref.~\onlinecite{Koide96}) 
and its radius (0.69 \AA) is close to
that of Cu$^{2+}$ ion (0.73 \AA).
Single-crystals grown by the floating-zone method contain almost the 
same Ni concentration as the starting powder materials.
It has been reported that the AF transition takes place
in polycrystalline 4\% Ni-doped CuGeO${}_3$
by means of specific heat measurements,\cite{ref3}
in single-crystal Cu${}_{0.967}$Ni${}_{0.033}$GeO${}_3$ 
by susceptibility measurement \cite{Koide96} and 1.5\% $\sim$ 6.0\% 
Ni-doped CuGeO$_3$
using neutron scattering techniques.\cite{ref5,Coad,Coad2}
Impurity concentration ($x$) dependences of $T_{SP}$ and 
$T_N$ of Cu${}_{1-x}$Ni$_{x}$GeO${}_3$ have not yet been determined.

In this paper we will study the temperature-vs-$x$ phase diagram of 
single-crystal Cu${}_{1-x}$Ni$_{x}$GeO${}_3$ in detail.
The disappearance of $T_{SP}$ and the corresponding increase 
of the $T_N$ at $x$ $\sim$ 0.020, 
which indicate the first-order phase transition between the D-AF and 
U-AF phases, are observed.
We also found that with decreasing $x$ the easy axis of 
Cu${}_{1-x}$Ni$_{x}$GeO${}_3$ in 
the D-AF phase changes from nearly the $a$ axis near $x_c$ to 
the direction containing the $a$ and $c$ components.
%This fact suggests a possibility of another phase transition to 
%an oblique antiferromagnetic (OAF) state in the D-AF state.

Single crystals of pure CuGeO$_3$ and 
Cu${}_{1-x}$Ni$_{x}$GeO${}_3$ with $0.005(1) \le x \le 0.060(2)$
were prepared by the floating-zone method.
The concentrations $x$ of Ni were determined
by electron-probe microanalysis (EPMA). 
Unexpected impurities or structural change with $x$ were not observed 
by means of x-ray diffraction after pulverization of the single crystals 
at room temperature.
The $a$-axis lattice parameter measured in single-crystal samples 
decreases monotonically as $x$ increases.
This fact supports that impurity ions are 
truly substituted in this system.

The magnetization measurements of the prepared samples 
were carried out using SQUID 
%(Superconducting QUantum Interference Device) 
magnetometer from 2 to 300 K.
Below 2 K, we measured the ac susceptibility of 0.5\% Ni-doped samples
using the $^3$He-$^4$He dilution refrigerator down to 20 mK.
Experimental details are described elsewhere.\cite{Manabe}

\begin{figure}[t]
\begin{center}
\includegraphics*[width=8cm]{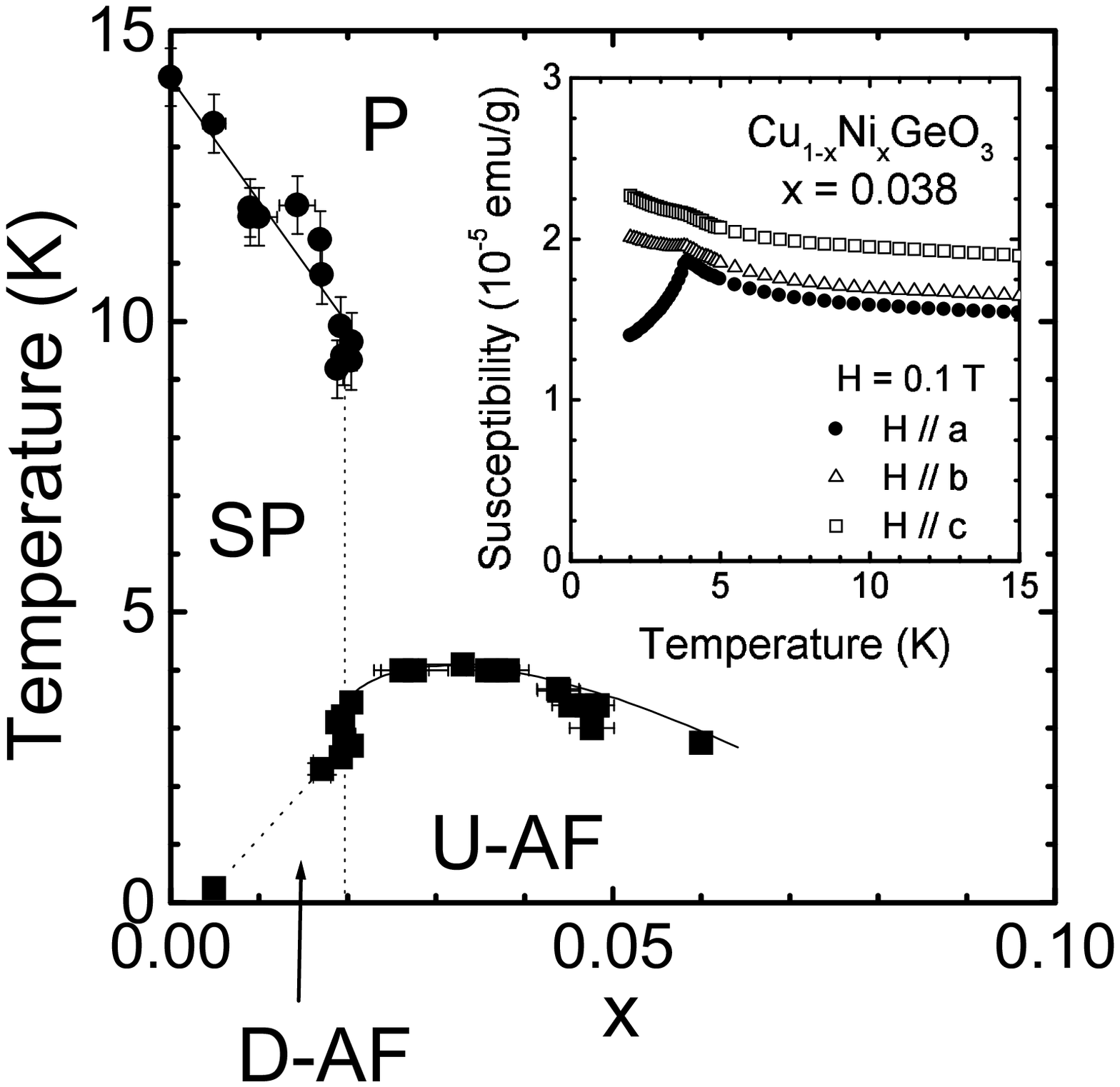}
%\vspace*{57truemm}
\end{center}
\caption
{The $x$ dependence of $T_{SP}$ (closed circles) 
and $T_{N}$ (closed squares) of single-crystal 
Cu${}_{1-x}$Ni${}_{x}$GeO${}_3$.
Paramagnetic, spin-Peierls, dimerized-antiferromagnetic, and
uniform-antiferromagnetic phases are
abbreviated as P, SP, D-AF and U-AF, respectively. 
The inset shows 
the temperature dependence of $\chi(T)$'s below 15 K 
under $H$ = 0.1 T of single-crystal Cu${}_{0.962}$Ni${}_{0.038}$GeO${}_3$.
}
\label{figure1}
\end{figure}

At first we show the temperature dependence of the susceptibility
along the $a$, $b$ and $c$ axes ($\chi_a(T)$, $\chi_b(T)$ and $\chi_c(T)$)
of 3.8\% Ni-doped CuGeO$_3$ under $H$ = 0.1 T below 15 K in the inset 
of Fig.~\ref{figure1}.
$\chi_a(T)$ has a cusp around 4.0 K and the magnitude of $\chi_a(T)$
decreases below 4.0 K, while the cusp around 4.0 K is less evident 
in $\chi_b(T)$ and $\chi_c(T)$. 
The magnetic field dependence of the magnetization in three directions 
($M_a(H)$, $M_b(H)$ and $M_c(H)$) were also measured and 
only $M_a(H)$ changes rapidly around 1.1 T, 
i.e., the spin-flop transition occurs (not shown here.\cite{full})
These behaviors indicate that the AF-LRO appears
below the transition temperature $T_N$ = 4.0 K.
However the easy axis is nearly along the $a$ 
axis,\cite{Coad} but all the spins do not align exactly along
the $a$ axis, because $\chi_a(T)$ does not 
tend to zero as $T\rightarrow 0$ K and moreover
$\chi_a(T$$\rightarrow$0 K$)$ 
has a significant value (see the inset of Fig.~\ref{figure1}).
On the contrary, in Zn- or Mg-doped CuGeO${}_3$ 
$\chi_c(T)$ tends to zero as $T\rightarrow 0$ K and therefore 
the easy axis is exactly along the $c$ axis.\cite{ref4,masuda}
This difference of spin orderings between Zn(Mg)- and Ni-doped
CuGeO${}_3$ is caused by that Ni${}^{2+}$ ions carry spin $S = 1$
and accordingly have the single-ion anisotropy, %
while Zn${}^{2+}$ (Mg${}^{2+}$) ions are nonmagnetic,
as will be discussed later in more detail.

Next, we show the $T$ - $x$ phase diagram of single-crystal
Cu${}_{1-x}$Ni${}_{x}$GeO${}_3$ in Fig.~\ref{figure1}.
We determined $T_{SP}$ as the temperature of the kink of $\chi (T)$
and $T_N$ as the temperature of the cusp of $\chi_a(T)$
with roughly estimated errors.
Figure~\ref{figure1} shows 
(a) the disappearance of the SP transition and 
(b) the abrupt increase of the $T_{N}$ from 2.5 to 3.4 K at 
nearly the same $x\sim$ 0.020.
We define this critical concentration as $x_c$.
These behaviors indicate the first-order phase transition between the
D-AF and U-AF phases, as was discovered in Mg-doped 
systems.\cite{masuda}
We also observed the broadening of the peaks of $\chi_a(T)$'s at $x\sim x_c$,
while the peaks are very sharp in both $x < x_c$ and $x > x_c$. 
This is due to the coexistence of two phases
near the first-order phase boundary,
as is always the case with a first-order phase transition.

\begin{figure}
%\vspace*{5mm}
\begin{center}
\includegraphics*[width=6.0cm]{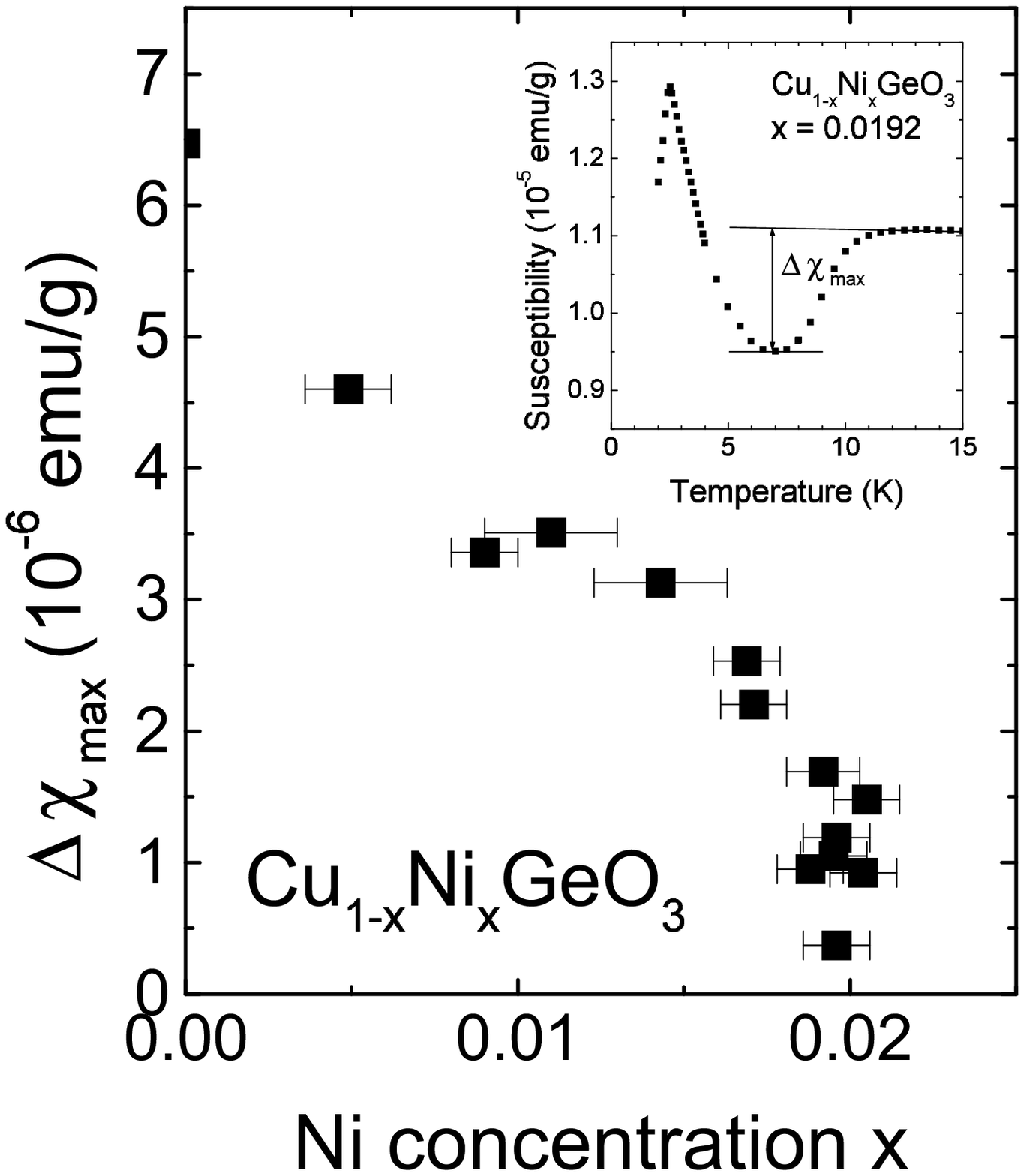}
%\vspace*{57truemm}
\end{center}
\caption
{The $x$ dependence of $\Delta \chi_{max}$ of single-crystal samples of  
Cu$_{1-x}$Ni$_x$GeO${}_3$.
The inset shows the method of the determination of 
$\Delta \chi_{max}$. }
\label{figure2}
\end{figure}

To clarify the existence of the first-order phase transition,
we show the $x$ dependence of $\Delta \chi_{max}$ in Fig.~\ref{figure2}.
Here $\Delta \chi_{max}$ 
is defined as the maximum of the difference between the 
linear extrapolation to $T<T_{SP}$ from $\chi_a(T)$ ($T>T_{SP}$) 
and $\chi_a(T)$ (see the inset of Fig.~\ref{figure2}).
$\Delta \chi_{max}$ can be taken as a measure of the magnitude of the 
dimerization.
A steep  decrease of $\Delta \chi_{max}$ around $x \sim$ 0.02 
can be explained by the sudden disappearance of $T_{SP}$ at $x_c$.
Near $x_c$ the SP and paramagnetic phases coexist because of the first-order
nature of the phase boundary at the vertical line at $x_c$ in the phase 
diagram of Fig.~\ref{figure1}, which is the reason why $\Delta\chi_{max}$ 
changes almost vertically at $x_c$ in Fig.~\ref{figure2}.
This is consistent with the existence of the first-order phase transition 
between the D-AF and U-AF phases.

\begin{figure}[t]
\begin{center}
\includegraphics[width=5cm]{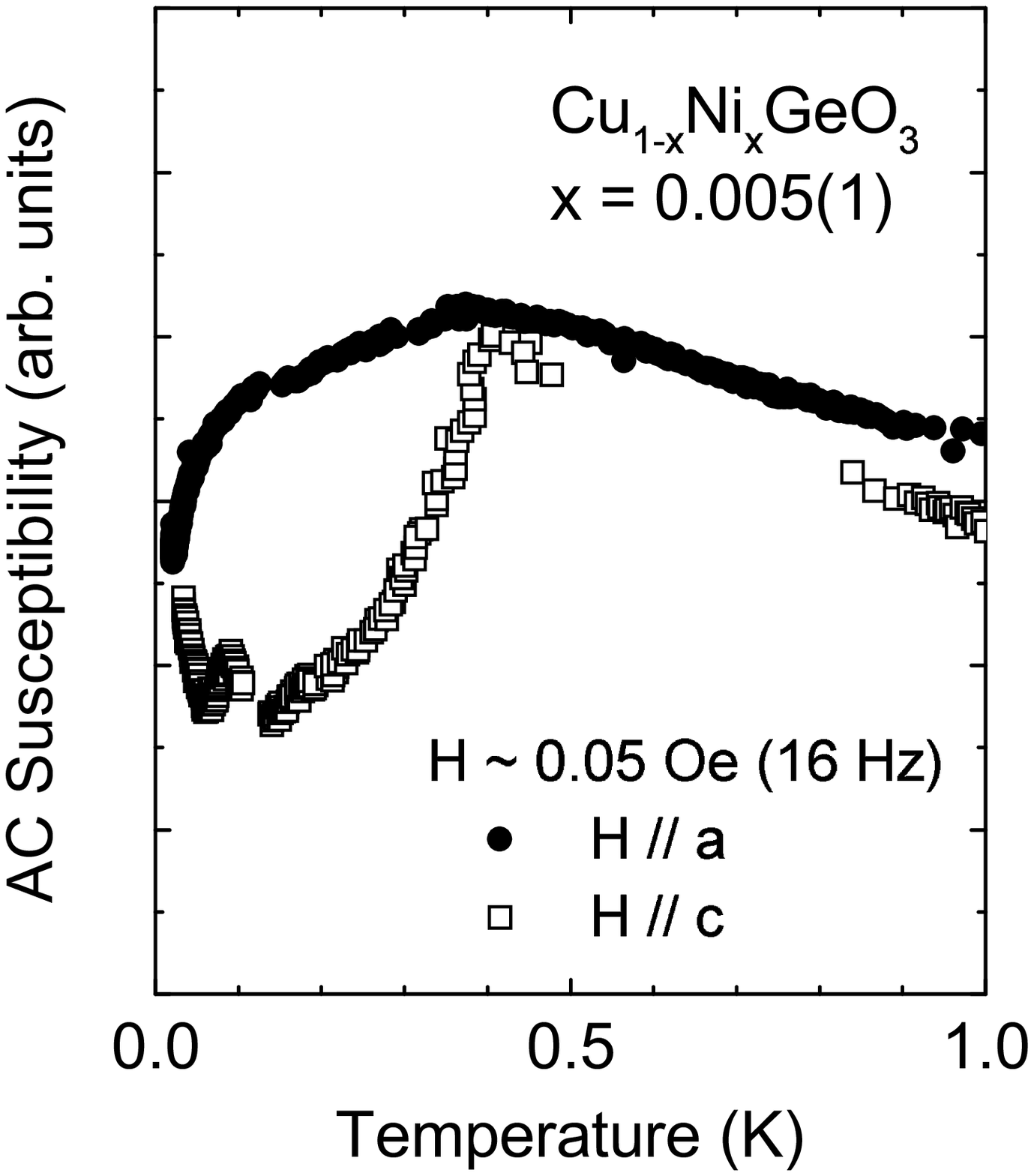}
%\vspace*{57truemm}
\end{center}
\caption
{
The temperature dependence of the magnetic susceptibility down to 20 mK 
of single-crystal Cu${}_{0.995}$Ni${}_{0.005}$GeO${}_3$.
%Closed circles and open squares indicate 
%$\chi_a(T)$ and $\chi_c(T)$, respectively.
}
\label{figure3}
\end{figure}

To study the anisotropy of the D-AF phase of 
Cu$_{1-x}$Ni$_x$GeO${}_3$ in detail,
we measured ac susceptibilities $\chi_a(T)$ and $\chi_c(T)$ 
of 0.5\% Ni-doped one 
down to 20 mK (Fig.~\ref{figure3}).
These measurements were carried out under ac magnetic field 
($\sim$ 0.05 Oe, 
16 Hz) applied parallel to the $a$ and $c$ axes.
Unfortunately the absolute values of the ac susceptibilities were not able
to be determined, but the AF transition was clearly observed.
In this sample, the SP and AF transitions occurred at 13.4 K and 
near 0.4 K, respectively.
We can see that $\chi_c(T)$ has a cusp around $T_{N}$ $\sim$ 0.4 K and 
the magnitude of $\chi_c(T)$ decreases below $T_N$, 
while the cusp around 0.4 K is less evident in $\chi_a(T)$. 
This indicates that the sublattice magnetizations of 
Cu${}_{0.995}$Ni${}_{0.005}$GeO${}_3$ 
have the components both along the $a$ and along the $c$ axes.

In Cu${}_{0.983}$Ni${}_{0.017}$GeO${}_3$ 
(which has the D-AF phase but $x\sim x_c$)
only $\chi_a(T)$ shows a cusp at $T_N$.
This definitely shows that the samples with $x\sim x_c$ in the D-AF phase 
have the easy axis nearly along the $a$ axis.
Combining these two experimental results we may think that another 
phase transition occurs in the D-AF state.
One of the phases is a D-AF phase with the easy axis 
nearly along the $a$ axis when
the Ni concentration $x$ is relatively high.
This is caused by the single-ion anisotropy of the $S=1$ spins 
on Ni${}^{2+}$ ions and by sufficiently strong 
exchange interaction between $S=1$ spins on Ni${}^{2+}$ ions and 
neighboring $S=1/2$ spins on Cu${}^{2+}$ ions.
The other is a D-AF state with the easy axis in the $a-c$ plane.
This can be compared with the oblique antiferromagnetic 
(OAF) phase, which appears in the mixture of two anisotropic AF states 
with perpendicular easy axes.\cite{Matsubara77}

In Fig.~\ref{figure3} 
we observe that $\chi_a(T)$ decreases and $\chi_c(T)$ increases as 
temperature decreases below 0.1 K.
This behavior is reproducible and is not able to be explained by the 
experimental errors.
It is one of the interesting problems 
to understand the reasons of these phenomena at ultra-low temperatures.

We have established that there occurs a first-order phase transition 
between the U-AF and D-AF phases in Ni-doped  CuGeO$_3$.
This transition occurs without the change of the direction of the 
easy axis. The easy axis remains nearly along the $a$ axis 
during the transition.
Recently Coad {\it et al.}\cite{Coad2} reported that the SP gap of 
3.2\% Ni-doped CuGeO$_3$ 
had collapsed almost completely, 
while that of 1.7\% sample had coexisted with the AF order.
Their results are consistent with our observations.

One of the most important issues in this Letter is whether the first-order 
phase transition between the D-AF and U-AF phases in Mg-doped CuGeO${}_3$ 
found by Masuda {\it et al.} \cite{masuda} is universal or not.
Our experimental results tell us that the same phenomenon occurs definitely in Ni-doped CuGeO${}_3$.
In Ni-doped CuGeO${}_3$ this phenomenon occurs 
without the change of the direction of the easy axis.
Considering the fact that Mg${}^{2+}$ ion is nonmagnetic and 
Ni${}^{2+}$ ion has $S=1$ spin on it, we may infer that this 
type of the first-order phase transition is universal at least 
for the substitution for Cu${}^{2+}$ ion.
Of course we must still study the detailed phase diagrams by using other 
dopants, e.g., Zn and Mn, which is being done at present.
A remaining problem is whether or not this transition occurs for the 
substitution  for Ge${}^{4+}$ ion by, e.g., Si${}^{4+}$ ion.
Recent report of the phase diagram of 
CuGe${}_{1-x}$Si${}_x$O${}_3$ (see Fig.~8 of Ref.~\onlinecite{Grenier98})
may indicate that this does not occur for Si-doping.
However, the temperature dependence of the magnetization of 
CuGe${}_{1-x}$Si${}_x$O${}_3$ shown in Fig.~5 of 
Ref.~\onlinecite{Grenier98} has broadening of the AF transition 
for $0.005\le x\le 0.024$ ($T_{SP}$ was observed  $x\le 0.008$.), 
in contrast to relatively sharp AF transition at $x=0.002$ and  0.05.
This is somewhat similar to that observed in Mg-doped 
CuGeO${}_3$ (see Fig.~3 of Ref.~\onlinecite{masuda}),
which indicated the coexistence of two phases (D-AF and U-AF phases)  
owing to a first-order phase transition.\cite{masuda}
Therefore more detailed study on the phase diagram of 
this system is needed to reach a definite conclusion. 

Here we will compare the phase diagrams of Ni-doped and Mg-doped 
(or Zn-doped) CuGeO${}_3$ in more detail.
In the U-AF phase and especially far from the phase boundary with the 
D-AF phase
the phase boundary with the paramagnetic phase is almost the same
in the two cases (see Fig.~2 of Ref.~\onlinecite{masuda}).
%The maximum $T_N$ is about 4 K and $T_N$ decreases with 
%increasing $x$ in almost the same manner.
This may be explained as follows.
The U-AF phase is supposed to be a conventional AF phase 
except for the existence of the disorder.
Every spins either on Cu${}^{2+}$ or on Ni${}^{2+}$ ions have effective 
magnetic moments $\mu_{eff}$ reduced by the quantum 
fluctuation.\cite{Martin}
Moreover in this concentration region the easy axes do not 
change in both cases, although the directions are different.
Therefore except for the existence of the $S=1$ spins on Ni${}^{2+}$ 
ions and the difference of the easy axes, both of the U-AF phases 
can be treated as classical N\'eel states, which are expected to 
be described by the mean-field theory.
This may be the reason why both of the U-AF phases behave similarly 
despite the difference of the easy axis.
To explain the decrease of $T_N$ with increasing $x$, we must consider 
the effect of the disorder, which plays an important role 
in AF materials into which the disorder 
is introduced.\cite{Uchinokura95}

Next we shall consider the competing effect of the anisotropies.
In Mg-doped (also Zn-doped) CuGeO${}_3$ the direction of the easy 
axis along the $c$ axis can be only explained by the small anisotropy
of the exchange interaction between the spins on Cu${}^{2+}$ ions,
because the spins have $S=1/2$ and do not have single-ion anisotropy
and because Zn${}^{2+}$ and Mg${}^{2+}$ ions are nonmagnetic.
On the other hand, spins on Ni${}^{2+}$ ions have $S=1$ and single-ion 
anisotropy.
When $x$ is large enough, $S=1$ spins on 
Ni${}^{2+}$ drive the whole $S=1/2$ spins on Cu${}^{2+}$ ions to orient 
nearly along the $a$ axis. 
This can be attained by sufficiently strong 
exchange interaction between $S=1$ spins on Ni${}^{2+}$ ions and 
neighboring $S=1/2$ spins on Cu${}^{2+}$ ions.
This explains the spin configurations of Mg- and Ni-doped CuGeO${}_3$
at large $x$.

In very low Zn concentration region of Cu${}_{1-x}$Zn${}_x$GeO${}_3$  
Manabe {\it et al.} experimentally showed that there is no critical 
concentration for the occurrence of AF-LRO and
also showed that the easy axis remains parallel to the 
$c$ axis down to $x=1.12(2)\times 10^{-3}$.\cite{Manabe}
The latter is easily understandable because there is only one kind
of anisotropy in Zn-(or Mg-)doped CuGeO${}_3$.
At low $x$ in Ni-doped CuGeO${}_3$, 
however, two kinds of the anisotropic energies
compete, because the effect of the single-ion anisotropy diminishes
with decreasing $x$.
Therefore at very low $x$ slightly 
anisotropic AF exchange interaction between the spins on Cu${}^{2+}$
ions dominates.
This is expected to drive the easy axis toward the $c$ axis
with decreasing $x$
and is the reason why we found the ``OAF state'' 
at $x\sim 0.005$ in Ni-doped CuGeO${}_3$.

The occurrence of the first-order phase transition between 
the D-AF and U-AF phases and its universality do not seem to have been 
theoretically explained conclusively.
Based on the phase Hamiltonian formulation applied to the 
explanation of the coexistence of dimerization and AF-LRO 
in Si-doped CuGeO${}_3$ at $T=0$ K,\cite{Fukuyama}
Saito proposes a model of a transition,  which, however, derives 
a second-order phase transition.\cite{Saito98}
Further considering the 3-dimensional interaction explicitly,
she claims that the first-order transition could be 
obtained.\cite{Saito-pc}
To construct a theory we must consider the fact that this transition
is universal for the substitution for Cu${}^{2+}$.
It also depends on whether this transition occurs for the substitution
for Ge${}^{4+}$ or not.

 In summary we have found that the first-order phase transition
observed in Mg-doped CuGeO${}_3$ occurs also in Ni-doped CuGeO${}_3$
with the different directions of the easy axis.
From this we infer that this transition occurs universally for
the substitution for Cu${}^{2+}$.
Competing anisotropies cause more complex phase diagram
in Ni-doped CuGeO${}_3$ than in Mg-
or Zn-doped CuGeO${}_3$ in the low concentration region.

We would like to thank I. Tsukada for valuable discussion and 
A. Sawa and H. Obara for the susceptibility 
measurements of some of the samples between 2 and 4.2 K.
This work was supported in part by Grant-in-Aid for Scientific 
Research (A),
Grant-in-Aid for Scientific Research on Priority Area 
``Mott Transition'', 
Grant-in-Aid for COE Research 
``Spin-Charge-Photon (SCP) Coupled System''
from the Ministry of Education, Science, Sports, and Culture,
and NEDO International Joint Research Grant.

\end{multicols}
\end{document}